\documentclass[aps,prd,showpacs,amssymb,twocolumn,
nofootinbib,superscriptaddress]{revtex4}
\usepackage{graphicx,hyperref}
\begin{document}

\title{Quantum time uncertainty in
Schwarzschild-anti-de Sitter black holes}

\author{Pablo
\surname{Gal\'an}}\email{galan@iem.cfmac.csic.es}
\affiliation{Instituto de Estructura de la Materia,
CSIC, Serrano 121, 28006 Madrid, Spain}

\author{
Luis J. \surname{Garay}}\email{luis.garay@fis.ucm.es}
\affiliation{Departamento de F\'isica Te\'orica II,
Universidad Complutense de Madrid, 28040 Madrid,
Spain} \affiliation{Instituto de Estructura de la
Materia, CSIC, Serrano 121, 28006 Madrid, Spain}

\author{Guillermo A. \surname{Mena Marug\'an}}
\email{mena@iem.cfmac.csic.es}
\affiliation{Instituto de Estructura de la Materia,
CSIC, Serrano 121, 28006 Madrid, Spain}

\date{\today}

\begin{abstract}

The combined action of gravity and quantum mechanics
gives rise to a minimum time uncertainty in the lowest
order approximation of a perturbative scheme, in which
quantum effects are regarded as corrections to the
classical spacetime geometry. From the nonperturbative
point of view, both gravity and quantum mechanics are
treated on equal footing in a description that already
contains all possible backreaction effects as those
above in a nonlinear manner. In this paper, the
existence or not of such minimum time uncertainty is
analyzed in the context of Schwarzschild-anti-de
Sitter black holes using the isolated horizon
formalism. We show that from a perturbative point of
view, a nonzero time uncertainty is generically
present owing to the energy scale introduced by the
cosmological constant, while in a quantization scheme
that includes nonperturbatively the effects of that
scale, an arbitrarily high time resolution can be
reached.

\end{abstract}


\maketitle
\renewcommand{\theequation}{\arabic{section}.\arabic{equation}}

\section{Introduction}

The emergence of a minimum time (or length)
uncertainty in the presence of gravity is often
analyzed by adopting perturbative approaches
\cite{Luis}. In standard quantum mechanics, the time
evolution is subject to the fourth Heisenberg
uncertainty relation: the time necessary to ensure
that a system has evolved is larger than the inverse
of the root mean square (rms) deviation of the energy
distribution. Therefore, to improve the time
sensitivity, one must consider states with an
increasing energy uncertainty, reaching a perfect time
resolution only if the energy is completely unknown.
In general relativity, on the other hand, Einstein
equations imply that an uncertainty in the energy of
the system causes uncertainty in the geometry of
spacetime and consequently in the measurement of time.
In this way the uncertainty in the time evolution of a
physical system is due to competing contributions from
both quantum and gravitational origin: energy
uncertainty must be large so that the quantum time
uncertainty is small; energy uncertainty must be small
so that spacetime geometry is not seriously disturbed.
As a result, an infinite time resolution seems
impossible in scenarios involving quantum and
gravitational effects and, indeed, several arguments
indicate that time uncertainty cannot be made as small
as desired in this kind of perturbative approaches (at
least in the next-to-leading-order approximation)
\cite{Luis,Pad,unc,unc2}. However, it is far from
clear whether this conclusion signals a fundamental
phenomenon in quantum gravity or can be eluded by
adopting a nonperturbative quantum description of the
gravitational processes, a possibility that the models
described in Refs. \cite{BMV,PG,PG2} actually suggest.

In those references it has been shown that some
systems present in fact a nonvanishing time
uncertainty when treated perturbatively, while the
time resolution may become arbitrarily high if
nonperturbative approaches are employed. It is
remarkable that  these fundamentally different
behaviors show up in quantum systems that are
sufficiently simple so as to allow full quantum
perturbative and nonperturbative descriptions. Whether
this is indeed a general feature of quantum
gravitational systems or not still remains to be seen.
Among these studies, we can find very different
systems such as Einstein-Rosen waves \cite{BMV} or the
so-called doubly special relativity formalisms
\cite{PG}. They share a common feature which is the
existence of modified dispersion relations, in the
sense that (nonlinear) redefinitions of the energy and
also of momentum reveal themselves as natural tools to
study these systems. Whether these redefinitions and
modified dispersion relations are just tools or
intrinsic characteristics of the system depends on the
point of view ---perturbative or not--- that one is
willing to adopt.

Black hole physics presents itself  as a suitable
scenario in which the possible appearance of a bound
to time resolution can be tested. These ideas are
further motivated by the existence of results that
hint at the discreteness of the spectrum of geometric
operators such as the area of the black hole horizon
\cite{area}. Although the discrete nature of these
operators does not necessarily imply a minimum
spacetime uncertainty, it certainly leads to a
spacetime picture which is not continuous at small
scales. A closely related issue which suggests the use
of black holes to analyze the perturbative nature (or
not) of minimum time uncertainties is that they
saturate the amount of information that can be stored
in a spacetime region of the same size \cite{Lloyd},
which, together with the ideas of the holographic
principle, provides a nontrivial time-uncertainty
lower bound \cite{Ng}.

In black hole physics, the definition of a notion of
horizon energy generically rests on normalization
conventions or global properties. As a consequence,
the energy does not have a genuine unambiguous
meaning, even when the horizon is associated with a
global Killing field. For instance, in the simple case
of event horizons in static and asymptotically flat
spacetimes, the energy can be determined using
conditions imposed in the asymptotic region, namely
that the global Killing field be normalized to the
unity there. In this way one assigns the (numerical
value of the) ADM energy to the horizon
\cite{wald,IH}. This type of criteria loose their
relevance when horizons are analyzed quasi-locally, so
that the knowledge of the whole spacetime is not
presumed \cite{IH,Ashtekar:2000hw}.

For the sake of simplicity, we will concentrate on the
nonrotating case in absence of matter fields. Then,
the energy of the horizon generates ``on shell'' time
translations on the horizon along a suitable vector
field which coincides with a null normal. This normal
can be freely rescaled by a constant, which may even
vary from one black hole solution to another. As we
have commented, if the spacetime is static and
asymptotically flat, one can eliminate this freedom by
restricting oneself precisely to the null normal
provided by the Killing field which has unit norm at
infinity. This is not possible in general (for
instance if there exists radiation in the exterior
region or the metric fails to have an appropriate
asymptotic symmetry). In order to analyze this sort of
systems, one is led to consider the notion of isolated
horizons \cite{IH} which need not be Killing horizons
nor require the entire spacetime history, as event
horizons do. Isolated horizons are in equilibrium: no
matter or radiation falls through them and their area
remains constant. In this formalism, elaborated by
Ashtekar {\it et al.}, the energy of the isolated
horizon can be defined using a Hamiltonian language.
The problem of normalization of the time vector field
which becomes the null normal on the horizon and
defines its mass is still present. Indeed, there is an
infinite family of parameter-dependent timelike vector
fields each defining a consistent Hamiltonian
evolution and a horizon energy. The possible
generators of Hamiltonian transformations on the space
of solutions (the covariant phase space) can be
identified as acceptable elections of the horizon
energy \cite{Ashtekar:2000hw} or, with an alternate
terminology, of the horizon mass function $\mu$.

The normalization of the time vector field is
important because the surface gravity $\kappa$ depends
on it. Given a horizon whose area is $A$, the mass
function $\mu$ and the time vector field along which
evolution takes place are closely related. Indeed, it
can be shown that the evolution defined by some
timelike vector field is Hamiltonian if and only if
the first law of black hole dynamics holds, $\delta
\mu=(\kappa/8\pi)\delta A$
\cite{Ashtekar:2000hw,note0}. But this still leaves a
lot of freedom in the choice of normalization of the
time vector and the mass function: any integrable
function $\kappa(A)$ leads to an acceptable mass
function $\mu(A)$ for the horizon.

In certain situations, the existence of different
allowed mass functions can be thought of as the result
of a modified normalization of the time vector that
incorporates gravitational effects with respect to a
background (e.g. a Schwarzschild spacetime). In
particular, here we will focus our attention on the
choice of mass function for black holes in a spacetime
with a negative cosmological constant and no matter
content ---we will refer to them generically as
Schwarzschild-anti-de Sitter (or Schwarzschild-AdS)
black holes. We will study whether the presence of a
scale fixed by the cosmological constant implies the
emergence of a minimum time uncertainty when one
(erroneously) assumes an effective quantum description
corresponding to an asymptotically flat space as the
starting point.

In a perturbative approach to the treatment of the
cosmological constant, one would begin with a
Schwarzschild black hole and the corresponding horizon
mass defined through the conventional asymptotic
normalization of the time vector field. Then, one
would proceed to introduce the effect of the
cosmological constant, deforming hence the spacetime
geometry. This deformation would change the
normalization of the time vector field and,
subsequently, result in a change of the horizon mass
function. If the modifications are incorporated
perturbatively, one would generally obtain a series of
successive corrections.

Alternatively, one could adopt a nonperturbative
approach in which a nonlinear global redefinition of
the mass and time parameters encompassing all the
effects would be in order. With these premises, we
will consider the different possibilities of
describing the quantum evolution in terms of a
parameter that corresponds either to the time
naturally associated with a Schwarzschild background
or to the time corresponding to the Schwarzschild-AdS
black holes. The latter can be regarded as the natural
time (on the horizon, or globally for static solutions
as we will see) whose definition includes the effects
of the cosmological constant. In this sense, we will
refer to these two types of quantization as
perturbative and nonperturbative, respectively, given
the distinct philosophy in the use of background
structures.

The rest of the paper is organized as follows. In the
following section, we introduce the two different
choices for the time and mass function in
Schwarzschild-AdS black holes discussed above, and we
describe the nonlinear relations between them. In Sec.
III we discuss a quantum framework for the description
of these black holes and analyze the time uncertainty
when one adopts a perturbative scheme for the
treatment of the effects of the cosmological constant,
proving that this uncertainty cannot generally vanish.
In addition, we study the behavior of the time
uncertainty in sectors of physical states with small
and large horizon mass compared with the scale
provided by the cosmological constant. In Sec. IV we
show that the time uncertainty can be made to vanish
in a nonperturbative quantization, in contrast with
the perturbative approach. We present our conclusions
and some further discussion in Sec. V. The Appendix is
devoted to some technical issues concerning the
calculation of Laplace transforms.

\section{Mass functions for Schwarzschild-AdS
black holes}
\setcounter{equation}{0}

We consider spacetimes with no matter fields that
present a single nonrotating isolated horizon as their
``internal boundary'' \cite{Ashtekar:2000hw} and that,
in principle, may be asymptotically flat or anti-de
Sitter (AdS) at infinity, depending on whether we
specialize to Schwarzschild or Schwarzschild-AdS black
holes. In the latter case, we assume the existence of
a negative cosmological constant. Then, given the
surface gravity $\kappa$, which depends on the
normalization of the time vector identified with the
null normal of the horizon, the first law of black
hole dynamics determines the horizon energy as a
function of the horizon area $A$, as we mentioned in
the introduction. On the covariant phase space, the
horizon energy plays the role of the generator of time
evolution on the horizon. The total Hamiltonian is
composed of two terms: the horizon energy and the
contribution at infinity, which generates asymptotic
timelike translations \cite{Ashtekar:2000hw}.
Furthermore, if one restricts oneself to the sector of
static solutions, the considered timelike vector field
can be chosen equal to the global static Killing
field. In this case, the total Hamiltonian must vanish
for symmetry reasons, implying that the horizon energy
coincides (in value) with the generator of asymptotic
time translations \cite{Ashtekar:2000hw}. But since
there exists only one vacuum static solution for each
value of the black hole area $A$, the mass function
$\mu(A)$ for the black hole is totally determined
provided that there is a preferred choice of
normalization for the static Killing field at
infinity. Note also that the mass function obtained in
this way generates evolution on the horizon with
respect to a time that, for the static solutions,
coincides precisely with the normalized asymptotic
time.

In the asymptotically flat case, it is natural to go
to the rest frame of the Schwarzschild black hole and
normalize the static Killing field to be the unit at
infinity. The generator of time translations at
infinity is then the ADM mass \cite{wald}. In the
asymptotically AdS case, one can introduce a similar
choice of asymptotic time by imposing a standard
normalization on the canonical generators of the AdS
group \cite{AdS,desitter} and arrive to a conserved
energy that, in the (static) Schwarzschild-AdS
spacetimes, provides the mass parameter of the black
hole \cite{ashAdS}. But if we have no access to the
asymptotic regions and analyze the horizon
quasilocally, we have no reason to choose one or the
other of these normalizations without additional
information apart from the condition that the first
law be satisfied. We will call Schwarzschild and
Schwarzschild-AdS the mass functions derived with
these two different normalizations for obvious
reasons. One of the aims of this paper is to explore
the consequences of using a Schwarzschild mass
function and its associated time parameter instead of
their Schwarzschild-AdS counterparts even when a
negative cosmological constant is present.

For completeness, let us recall that the static metric
of Schwarzschild and Schwarzschild-AdS spacetimes can
be respectively expressed in the well known form
\begin{eqnarray}
ds^2&=&-\left[1-\frac{2M}{r}\right]dT^2+
\left[1-\frac{2M}{r}\right]^{-1}dr^2+r^2d\Omega^2,
\nonumber\\
\label{Schw-AdS}
ds^2&=&-\left[1-\frac{2m}{r}+\frac{\Lambda}{3}r^2\right]
dt^2+\left[1-\frac{2m}{r}+\frac{\Lambda}{3}r^2\right]^{-1}
\nonumber\\&\times& dr^2 +r^2d\Omega^2,
\end{eqnarray}
where $d\Omega^2$ is the metric on the unit
two-sphere, $\Lambda>0$ is the absolute value of the
negative cosmological constant, and $M$ and $m$ are
the mass parameters of the respective black hole
solutions. These mass parameters coincide in numerical
value with the corresponding generators of asymptotic
time translations (in $T$ and $t$, respectively) at
infinity. For instance, in the asymptotically flat
case, $M$ is the ADM mass of the Schwarzschild black
hole. The horizon area is $A=4\pi r_h^2$ in both
cases, where $r_h$ is given by the positive zero of
the diagonal time component of the metric.

For a Schwarzschild and a Schwarzschild-AdS black
hole, respectively, the horizon mass $\mu(A)$ is hence
\begin{eqnarray}
 M(A)&=&
\sqrt{\frac{A}{16\pi}},\nonumber\\
 m(A)&=&
\sqrt{\frac{A}{16\pi}}+\frac{4}{3}\Lambda
\left(\sqrt{\frac{A}{16\pi}}\right)^{3}.\label{22}
\end{eqnarray}
The notation $M(A)$ and $m(A)$ emphasizes the fact
that the numerical value of these functions coincide
with the mass parameters $M$ and $m$ of the
corresponding static solutions. In the following, we
will not display explicitly this area dependence. The
relation between the two considered mass functions is
clearly
\begin{equation}\label{g}
m=M+\frac{4}{3}\Lambda M^3.
\end{equation}
It is worth remarking that $m$ and $M$ indeed coincide in the limit
$\Lambda\rightarrow 0$. Abusing of the notation we will call $T$ and
$t$ the times which parameterize the evolution generated by the mass
functions $M$ and $m$ on the horizon, respectively. Remember that
these times parameterize also the asymptotic time translations (with
a convenient convention of signs) if one restricts oneself to static
solutions, and that the normalization adopted at infinity would be
the standard one for $T$ if the spacetime were asymptotically flat,
while it is standard for $t$ if the asymptotic behavior is in fact
AdS.

Let us then assume from now on that we are studying
the horizon of a Schwarzschild-AdS black hole
(quasilocally). We will refer to the times $t$ and $T$
as the physical and auxiliary ones, respectively,
since they provide the AdS time and its counterpart
for vanishing cosmological constant in the static
sector, as we have commented. These times differ in a
(solution-dependent, i.e. mass-dependent) constant
factor which relates the two normalizations under
consideration:
\begin{equation}\label{t}
t=V(M;\Lambda)T.
\end{equation}
Obviously, we have chosen the same origin for both
times, so that they vanish simultaneously. The surface
gravities associated with these different
normalizations of the time vector field are related by
$\kappa_t=V^{-1}\kappa_T$. On the other hand, the
first law imposes that the horizon mass in each case
is determined by
\begin{equation}
\delta m=\frac{\kappa_t}{8\pi}\delta A,\qquad \delta
M=\frac{\kappa_T}{8\pi}\delta A.
\end{equation}
Dividing both expressions, we obtain the factor $V$:
\begin{equation}\label{V}
V(M;\Lambda)=\left[\frac{\partial m}{\partial M}
\right]^{-1}=\frac{1}{1+4\Lambda M^2}.
\end{equation}
Note that this factor is strictly positive, so that
the time relation (\ref{t}) is a bijection from
$\mathbb{R}$ to $\mathbb{R}$. Besides, $V$ becomes the
unit when $\Lambda$ vanishes, so that the two analyzed
times coincide in the limit of vanishing cosmological
constant. Finally, notice that $V$ is not a polynomial
in $\Lambda$, so that its Taylor expansion around
$\Lambda=0$ would provide a (perturbative) series with
an infinite number of terms.

\section{Time Uncertainty: Perturbative Case}
\setcounter{equation}{0}

We will now assume the existence of a quantum
description for our covariant phase space
corresponding to solutions in a vacuum with an
internal nonrotating isolated horizon. We will not
adhere to a particular quantization to try to keep our
discussion as generic as possible, and rather base our
analysis on general features of the quantum theory. In
particular, we expect that the horizon area be
represented by a quantum observable (with positive
spectrum), since it is a well defined (positive)
quantity on the considered covariant phase space. This
is known to be the case, for instance, in loop quantum
gravity \cite{area,qubh}. Therefore, also the
Schwarzschild and Schwarzschild-AdS mass functions,
being functions of the horizon area, will be
represented by quantum observables, which can be
defined by means of the spectral theorem. Note also
that, even if there exists a cosmological constant and
the covariant phase space corresponds to
asymptotically AdS spacetimes, the Schwarzschild mass
function will still be well defined and what will
cease to be applicable is just its physical
interpretation in terms of the normalization of the
time vector field at infinity for the static
solutions.

In this section we will study the consequences of
adopting a quantum description in which the parameter
of the evolution at the horizon is the auxiliary time
$T$ (i.e. the choice of time that would correspond to
the standard asymptotic time for static solutions if
the cosmological constant vanished). This would be the
natural choice of time if one started the analysis
obviating the presence of a cosmological constant and
decided to incorporate it afterwards by perturbative
means. If the reduction to the static sector of the
covariant phase space is meaningful quantum
mechanically, the time $T$ would play the role of the
asymptotic time parameter in this reduced theory, but
with a nonstandard (mass-dependent) normalization
caused by a negligent account of the existence of a
cosmological constant \cite{quantum}. We leave to the
next section the discussion of the case in which the
evolution parameter is chosen to be the physical time
$t$. The remarkable point is that if the role of
evolution parameter is indeed assigned to the
auxiliary time, the physical time is represented as a
one-parameter family of quantum observables
\begin{equation}\label{tquan}
\hat{t}=\hat{V}T,
\end{equation}
where $\hat{V}$ can be constructed from the operator
representing the Schwarzschild mass function as
$\hat{V}:=V(\hat{M};\Lambda)$.

In order to calculate the uncertainty in the elapsed
physical time $\hat{t}$ (setting the starting time for
our observations at $t=T=0$), we employ the following
procedure. Given a quantum state, one can measure the
probability density of the operator $\hat V$. We call
$\Delta V$ and $\langle\hat{V}\rangle$ its rms
deviation and mean value, respectively. On the other
hand (as in standard quantum mechanics), the value of
the elapsed time $T$ can be deduced by analyzing the
evolution of the probability densities of observables
in the quantum state, with a resolution that cannot be
better, for each given observable, than the lapse of
time required for a change in its expectation value of
the amount of the rms deviation. Therefore, this
resolution is limited by the fourth Heisenberg
relation. Via the analysis of collections of
probability densities of observables, the above
process will lead to a statistical distribution for
the value of the parameter $T$ (regarded as a random
variable), that will be described by a probability
density $\rho(T).$ The corresponding mean value will
be denoted by $\overline{T}$. According to our
comments and recalling that the evolution in the
parameter $T$ is generated by $\hat M$, the
uncertainty $\Delta T$ of this distribution must
satisfy the fourth Heisenberg relation $\Delta T\Delta
M\geq 1/2$. A double average is hence involved in the
calculation of $\Delta t$ since we have to calculate
the quantum expectation value $\langle\cdot\rangle$
and also the statistical average over the value of
$T$:
\begin{eqnarray}\label{unc}
(\Delta t)^2&=& \int dT \rho(T) \;\langle\;
T^2\hat{V}^2-\overline{T}^2
\langle\hat{V}\rangle^2\;\rangle \nonumber\\
&=&(\overline{T}\Delta V)^2+(\Delta T\Delta V)^2+
\langle\hat{V}\rangle^2(\Delta T)^2.
\end{eqnarray}
Because the last expression is a sum of positive terms, the time
uncertainty vanishes if and only if all of them are equal to zero.
We show now that this will not happen at any generic time $\overline
T\neq 0$ \cite{note}. For this purpose, it is convenient to express
Eq. (\ref{unc}) as
\begin{eqnarray}\label{unc2}
(\Delta t)^2&=&\overline{T}^2(\Delta
V)^2+\langle\hat{V}^2\rangle(\Delta T)^2
\nonumber\\
&\geq& \overline{T}^2(\Delta
V)^2+\frac{1}{4}\;\frac{\langle\hat{V}^2\rangle}{(\Delta
M)^2},
\end{eqnarray}
where we have used the fourth Heisenberg relation.

In order for the uncertainty $\Delta t$ to vanish, the
first term in the last inequality must be zero which,
for $\bar T\neq 0$, implies that $\Delta V$ must
vanish. But then the second term cannot vanish. To
prove this, notice that the function $V$ given by Eq.
(\ref{V}) is strictly monotonic in $M\in\mathbb{R}^+$,
so that it provides a one-to-one map. Thus, the
spectral theorem assures that the eigenstates of the
operators corresponding to $V$ and $M$ coincide, and
the condition $\Delta V=0$ implies that $\Delta M=0$.
On the other hand, since $V^2$ is a strictly positive
function of $M$, $\hat{V}^2$ is a positive operator
and hence $\langle\hat{V}^2\rangle$ does not vanish
\cite{nota}. Thus, if $\Delta V$ becomes zero, the
last term in inequality (\ref{unc2}) gets unboundedly
large. It is worth remarking that this result is
completely general: $\Delta t$ does not vanish
regardless of the choice of quantum state (provided
that \cite{nota} is taken into account).

In the rest of this section we will analyze in detail
the behavior of this nonvanishing time uncertainty in
two different regimes which are respectively included
in the sectors of quantum states with small and large
(expectation values of the) Schwarzschild mass in
comparison with the mass scale provided by the
cosmological constant. We will also relate this
behavior with suggested bounds for the time resolution
that have appeared in the literature.

\subsection{Small mass and holographic uncertainty}

For small Schwarzschild masses with respect to the
scale determined by the cosmological constant
(name\-ly for $4\Lambda M^2<1$), expression (\ref{V})
can be expanded as a Taylor series around the origin,
obtaining
\begin{equation}\label{per}
V=\frac{1}{1+4\Lambda M^2}=
\sum_{n=0}^{\infty}\left(-4\Lambda M^2\right)^n.
\end{equation}
This series can also be viewed as a perturbative
expansion of $V$ in terms of the cosmological constant
for any given finite mass $M$. A direct calculation,
taking the above series as the formal definition of
$V$, gives then
\begin{eqnarray}
\langle\hat{V}\rangle^2&=
&\sum_{n=0}^{\infty}(-4\Lambda)^n\sum_{l=0}^{n}
\langle\hat{M}^{2n-2l}\rangle\langle
\hat{M}^{2l}\rangle,\nonumber\\
\langle\hat{V}^2\rangle&=
&\sum_{n=0}^{\infty}(n+1)(-4\Lambda)^n\langle
\hat{M}^{2n}\rangle,\nonumber\\
(\Delta V)^2&=&\sum_{n=1}^{\infty}(-4\Lambda)^n
\left[n\langle\hat{M}^{2n}\rangle
\right.\nonumber\\&-&\left.
\sum_{l=1}^{n}\langle\hat{M}^{2n-2l}\rangle\langle
\hat{M}^{2l}\rangle\right].\label{uncv}
\end{eqnarray}
Introducing these expressions in Eq. (\ref{unc2}) we
get
\begin{eqnarray}\label{uncser}
(\Delta t)^2&\geq&\frac{1}{4(\Delta
M)^2}+\sum_{n=1}^{\infty}(-4\Lambda)^n\left[
\frac{n+1}{4}\frac{\langle\hat{M}^{2n}\rangle}
{(\Delta M)^2}\right.\nonumber
\\&+&\left.\overline{T}^2\!\left(n\langle\hat{M}^{2n}
\rangle-\sum_{l=1}^{n}\langle\hat{M}^{2n-2l}\rangle
\langle\hat{M}^{2l}\rangle\right)\right]\!.
\end{eqnarray}

Let us now restrict our discussion to the sector of
quantum states with sufficiently small Schwarzschild
mass, in the sense that
\begin{equation}\label{sta}
1\gg n \Lambda^{n}\langle\hat{M}^{2n}\rangle,\quad
\forall n\geq 1.
\end{equation}
Alternatively, one can regard our discussion as
corresponding to the limit of vanishing cosmological
constant in the sector of quantum states with bounded
even moments $\langle\hat{M}^{2n}\rangle$ for the
quantum observable $\hat{M}$ \cite{nota2}. Conditions
(\ref{sta}) are then satisfied when $\Lambda$ is
sufficiently small.

In these circumstances, one can neglect the
$\overline{T}$-in\-dependent corrections to the value
of the (square) time uncertainty for zero cosmological
constant, $1/[4(\Delta M)^2]$. In principle, the
$\overline{T}$-dependent corrections cannot be
neglected if the auxiliary time can become unboundedly
large, because $\overline{T}^2$ could compensate for
the smallness of the factors containing powers of
$\Lambda$ and the Schwarzschild mass. Nonetheless, we
should expect from Eq. (\ref{uncv}) that $\Delta
V\approx 4 \Lambda \Delta(M^2)$ in the spirit of our
small mass approximation. For instance, one can prove
that this is actually so when inequality (\ref{sta})
holds if $\langle\hat{M}^{4}\rangle$ is of the same
order or smaller than $[\Delta(M^2)]^2$ and besides
\begin{equation}\label{moresta}
[\Delta(M^2)]^2\gg n \Lambda^{n-2}
\langle\hat{M}^{2n}\rangle \quad \forall n>
2.
\end{equation}
We then reach the following (approximate) lower bound for the
uncertainty  $\Delta t$:
\begin{equation}\label{aproxunc}(\Delta t)^2\gtrsim
\frac{1}{4(\Delta M)^2}+16\Lambda^2[\Delta
(M^2)]^2\overline {T}^2.\end{equation} Again, this
equation can be interpreted as the Heisenberg bound
for the time uncertainty when $\Lambda$ vanishes
modified with the first significant perturbative
correction in the limit of negligibly small
cosmological constant \cite{note2}.

We next define
\begin{equation}\label{omega}
\omega:=\frac{\Delta (M^2)}{(\Delta M)^2}.
\end{equation}
Although $\omega$ and $\Delta M$ can be treated in
principle as independent parameters in the Hilbert
space of quantum states, inasmuch as they involve
different moments of the quantum probability
distribution for the observable $\hat{M}$ [$\Delta
(M^2)$ involves $\langle\hat{M}^{4}\rangle$ whereas
$\Delta M$ does not], one can argue that $\omega$,
which is nonnegative by construction, has to be
bounded from below by a strictly positive number.
Indeed, on the one hand, since $M^2$ is a monotonic
function of $M\in\mathbb{R}^+$, the spectral theorem
ensures that $\Delta (M^2)$ vanishes if and only if so
does $\Delta M$. In fact, it is possible to see that
$(\Delta M)^2$ approaches zero faster than $\Delta
(M^2)$. On the other hand, employing Eq. (\ref{sta})
one can show that, for each given value of $\Lambda$,
$\Delta M$ is bounded from above in the considered
sector of quantum states. Hence, $\omega$ cannot
approach zero by letting $\Delta M$ diverge. As a
result, one can convince oneself that there must exist
a (possibly $\Lambda$-dependent) positive number
$\omega_0>0$ such that $\omega\geq\omega_0$.

We then have
\begin{equation}(\Delta t)^2\gtrsim\frac{1}
{4(\Delta M)^2}+
16\Lambda^2\omega^2_0\overline{ T}^2(\Delta M)^4.
\end{equation}
This expression can be minimized with respect to its
dependence in $\Delta M$, obtaining in this way the
minimum time uncertainty at each instant of time
$\overline{T}$ in the sector of states under study.
The extremum at time $\overline{T}$ is reached on
states with $(\Delta
M)_{min}^{-6}=128\omega^2_0\Lambda^2\overline{T}^2$,
and the corresponding minimum uncertainty is
\begin{equation}
(\Delta t)_{min}= \sqrt{3}\left(\frac{\Lambda
\omega_0\overline{T}}{2}\right)^{1/3},\end{equation}
which scales with the auxiliary time as
$\overline{T}^{1/3}$.

It is worth remarking that this behavior of the time
uncertainty is precisely the same that one finds by
applying holographic arguments to black holes in
quantum gravity \cite{Ng}. Nevertheless, it is
important to point out that our result depends on the
type of quantum states that one chooses. We would find
a generally different auxiliary time dependence for
the time uncertainty if we considered states other
than those satisfying the conditions commented above,
which in particular imply that the even moments of the
Schwarzschild mass are sufficiently small compared
with the mass scale determined by the cosmological
constant.

\subsection{Large mass and linear uncertainty}

We now analyze the time uncertainty on quantum states
that present an approximate Gaussian distribution in
the Schwarzschild mass, peaked around a large
eigenvalue in Planck units. In principle, one might
worry about the very assumption of the existence of
such states, since we have preferred not to adhere to
any specific quantization of our covariant phase space
and therefore we cannot give precise statements about
the spectrum of the Schwarzschild mass. For instance,
from the discreteness of the area spectrum in loop
quantum gravity \cite{area}, one would expect that a
loop quantization of our system would lead to a
discrete Schwarzschild mass. Nonetheless, for large
values of the Schwarzschild mass function in Planck
units [and then of the AdS mass, see Eq. (\ref{g})],
one should expect the (semi-)classical description of
the black hole to be a fairly good approximation.
Hence, the work hypothesis that the mass spectrum is
almost continuous in the region of {\it macroscopic}
black holes is a reasonable supposition, when not a
requirement that should be imposed in order to select
the physically admissible quantizations. On the other
hand, we will also restrict the mass eigenvalue around
which the quantum state peaks to be at least of the
order of the mass scale provided by the cosmological
constant.

In more detail, we are going to concentrate our
discussion on quantum states whose associated
probability distribution for the Schwarzschild mass
has the approximate Gaussian form:
\begin{equation}\label{state}
\overline{\rho}(M)=\frac{1}{N}
e^{-(\frac{M}{\sigma}-\mu)^2}\end{equation} and allow
$M$ to run over the positive real axis. The factor $N$
is a normalization constant, so that $\overline{\rho}$
is a normalized distribution. It is straightforward to
see that
\begin{equation}N=\sigma
\int_{-\mu}^{\infty}du \,e^{-u^2}:=\sigma
\Xi(\mu).\end{equation} The function $\Xi(\mu)$ can be
expressed in terms of the common error function
\cite{grad}:
\begin{eqnarray}\Xi(\mu)&=&
\frac{\sqrt{\pi}}{2}[1+{\rm erf}(\mu)],\nonumber
\\
{\rm erf}(\mu)&=&\frac{2}{\sqrt{\pi}} \int_0^{\mu}du\,
e^{-u^2}.\end{eqnarray} Remember that ${\rm erf}(\mu)$
tends to the unit if $\mu\rightarrow\infty$. On the
other hand, $\mu$ is related with the mean value of
the mass distribution measured in $\sigma$-units.
Explicitly, \begin{equation} \langle \hat{M}\rangle=
\sigma\left[\mu+\frac{e^{-\mu^2}}{2\Xi(\mu)}\right].
\end{equation}
Hence, for large values of $\mu$, we have both that
the mean value of $\hat{M}$ is large and that it
coincides with $\sigma\mu$ up to negligible
exponentially decreasing terms. Finally $\sigma$,
apart from providing a unit of mass, would correspond
to the rms deviation of a usual Gaussian were the
variable $M$ defined on the entire real line, rather
than on the positive real axis. Taking into account
this restriction on the range of $M$, one can check
that
\begin{equation}(\Delta M)^2=\frac{\sigma^2}{2}
\left[1-\frac{\mu e^{-\mu^2}}{\Xi(\mu)}
-\frac{e^{-2\mu^2}}{2\Xi(\mu)^2}\right].
\end{equation}
Therefore, for large $\mu$ and up to exponentially
small corrections, the rms deviation becomes
$\sigma/\sqrt{2}$. In total, we see that the
distribution (\ref{state}) can be considered a well
peaked Gaussian around large values of the
Schwarzschild mass (compared with the Planck mass) if
$\mu\gg 1$ and $ \sigma$ is not considerably large in
Planck units.

With the probability distribution (\ref{state}), one
obtains for the operator $\hat{V}$
\begin{equation}\label{Vn}
\langle\hat{V}^n\rangle=\frac{1}
{\sqrt{\lambda}\Xi(\mu)} \int_{-\gamma}^{\infty}
dv\frac{e^{-v^2/\lambda}}{\left[1+(v+\gamma)^2
\right]^n}.\end{equation} We have defined
\begin{equation}
\lambda:=4\Lambda\sigma^2,\quad
\gamma:=\mu\sqrt{\lambda},\quad
v:=\sqrt{\lambda}\left[\frac{M}{\sigma}-\mu\right].
\end{equation}
Note that expression (\ref{Vn}) has the form of a
Laplace integral. In the Appendix we show how to
compute its asymptotic series expansions in powers of
$\lambda$ in the limit of vanishing cosmological
constant. In those computations, we treat $\gamma$ as
a given number. Up to a constant numerical factor,
this number is precisely the mean value of $\hat{M}$
expressed in terms of the mass scale provided by the
cosmological constant (in the approximation of large
$\mu$). Remember that we have assumed that these two
masses are of the same order, so that we cannot
neglect $\gamma$, even if we analyze the asymptotic
limit $\lambda\rightarrow 0$.

So, we now specialize the calculations to the region
of large black hole masses, where the consideration of
the approximate Gaussian distributions (\ref{state})
is justified, and study the situation of a very small
cosmological constant with fixed $\gamma$. Under these
circumstances, one can approximate the function
$\Xi(\mu)$ to $\sqrt{\pi}$ and $\Delta M$ to
$\sigma/\sqrt{2}$ by disregarding exponentially small
terms. In addition, one gets the following
leading-order contributions for the moments of
$\hat{V}$ (see the Appendix for details):
\begin{equation}
\langle\hat{V}^2\rangle\approx
\frac{1}{(1+\gamma^2)^2},\quad
(\Delta\hat{V})^2\approx\lambda\frac{2\gamma^2}
{(1+\gamma^2)^4}.
\end{equation}
With all these results, one obtains from Eq.
(\ref{unc2}):
\begin{equation}(\Delta t)^2
\gtrsim\frac{1}{2(1+\gamma^2)^2\sigma^2}+\lambda
\frac{2\gamma^2}{(1+\gamma^2)^4}\overline{T}^2.
\end{equation}
The last term on the right hand side has been
conserved, even though it is of higher-order in
$\lambda$ in comparison with the first one, because
the auxiliary time can be unboundedly large. In this
way, we obtain a time uncertainty that is always
bounded from below by a positive constant contribution
$1/[\sqrt{2}(1+\gamma^2)\sigma]$ and that for large
(auxiliary) times grows linearly with $\overline{T}$
[and thus with $\langle t \rangle$, see Eq.
(\ref{t})].

\section{Time Uncertainty: Nonperturbative Case}

We turn now to the discussion of the time uncertainty
when one considers that the quantum evolution on the
horizon is dictated by the Schwarzschild-AdS mass
function and the corresponding evolution parameter is
thus the physical time $t$. Remember that, if the
reduction to the static sector of the covariant phase
space is meaningful quantum mechanically, the physical
time can be viewed in the reduced theory as the
asymptotic AdS time parameter with the standard
normalization at infinity.

The fundamental difference with respect to our
analysis in the previous section is that now $t$ is
not a one-parameter family of observables, but instead
a genuine parameter. Its uncertainty is only limited
now by the fourth Heisenberg relation, namely $\Delta
t\Delta m\geq 1/2$.

As a consequence, in this nonperturbative quantum
description, the resolution for the physical time is
intrinsically bounded if and only if the same happens
for the AdS mass $m$. The conclusion does not depend
on other details of the system. The only relevant
point is whether the range of the horizon energy is
bounded or not. This range is determined in the
quantum theory by the spectrum of the observable
$\hat{m}:=m(\hat{A})$ obtained from Eq. (\ref{22}).
Since the classical function $m(A)$ tends to infinity
when so does $A$, the spectrum of the AdS mass will be
indeed unbounded if the same happens with the black
hole area spectrum. This last assumption is certainly
reasonable, since one should expect no important
quantum deviations from the (semi-)classical
description for {\it macroscopic} black holes with
large horizon areas (for instance, the area spectrum
is known to be unbounded in loop quantum gravity
\cite{area}). Therefore, a finite time resolution is
not a necessary consequence of the quantization of the
system. At least in this nonperturbative framework,
the quantum resolution in the physical time can be
made as large as desired.

\section{Conclusion}

We have argued that, for Schwarzschild-AdS black holes, a
perturbative approach to quantization gives rise to a (time and
quantum-state dependent) lower bound on the time uncertainty,
while nonperturbative treatments allow for arbitrarily high
time resolutions. This conclusion has been reached by analyzing
the quantization of a Schwarzschild-AdS black hole using
different mass functions for the horizon, each of them
generating evolution on the horizon with respect to a different
time parameter.

We started with the mass function that, for static
solutions, would correspond to Schwarzschild black
holes with a standard normalization at infinity of the
time vector field as if, instead of dealing with a
Schwarzschild-AdS black hole, we had an asymptotically
flat spacetime. The presence of the cosmological
constant was then introduced as a contribution to the
spacetime energy which modifies the normalization of
the time vector field and hence the mass function,
leading to effects that can be described by a
perturbative power series in the cosmological constant
[see e.g. Eq. (\ref{per})].

On the other hand a nonperturbative quantization has
also been carried out. In this approach, the mass
function for the horizon is that obtained with the
standard normalization of the time vector field at
infinity for the kind of spacetimes that we are truly
studying, i.e. Schwarzschild-AdS black holes. By
adopting this choice of the mass function, one is
naturally including the effects of the cosmological
constant in our description from the beginning. The
quantizations based on the two commented choices of
the mass function are not equivalent, because the
physical time, which reproduces the AdS time on static
solutions, is described in one case as a parameter,
while in the other case is represented as a
one-parameter family of observables. Indeed we have
seen that the two quantization schemes yield different
physical results.

In the nonperturbative approach, time uncertainty is
just dictated by the fourth Heisenberg uncertainty
principle and therefore presents no lower bound,
because the AdS mass is an unbounded function of the
horizon area and (it seems reasonable to assume that)
it is represented by an unbounded operator in the
quantum theory.

From the perturbative point of view, on the other
hand, we have seen that the physical time uncertainty
can never be made equal to zero. In the limit of a
negligibly small cosmological constant, this
uncertainty can be expanded in an asymptotic power
series. We have shown that the minimum time
uncertainty, although universally nonvanishing, has a
behavior that strongly depends on the quantum state.
Indeed we have considered different quantum states
which describe black holes with small and large mass
with respect to the mass scale supplied by the
cosmological constant. These lead to different types
of time dependence, which have the general form
$\Delta t_{min}\sim \langle t\rangle^{1-\delta}$ (in
Planck units) with $0\leq \delta \leq 1$, as discussed
for instance in Ref. \cite{Ng}. The specific classes
of quantum states that we have analyzed illustrate the
behaviors for $\delta=1,\ 2/3$, and $0$. These kinds
of uncertainties correspond, respectively, to the
cases of a constant (Planck) minimum time uncertainty,
a holographic-principle type of uncertainty, and a
constant relative time uncertainty (so that the
absolute one is cumulative and grows linearly with
time).

On the other hand, one might try to extend our
analysis to the case of Schwarzschild-de Sitter black
holes. A standard normalization of the time vector
\cite{desitter} would lead to a correspondence similar
to Eq. (\ref{g}), but with a flip of sign in the last
term, arising now from a positive cosmological
constant. As a consequence, the relation between the
two considered masses would be one-to-one only in a
bounded interval of positive masses including the
origin. This restriction on the allowed masses, that
would affect the analysis, is not casual: the upper
bound in the Schwarzschild-de Sitter mass is just the
value for which one gets a static solution with
coincident black hole and cosmological horizons.
Furthermore, the presence of a cosmological horizon
casts shadows on the naturalness of the standard
normalization adopted for the time vector
\cite{desitter} and, in any case, makes unavailable a
physical interpretation in terms of asymptotic
timelike translations on static solutions. Owing to
these reasons, we have restricted our discussion to
Schwarzschild-AdS black holes.

Finally, let us emphasize that the fact that a
nonvanishing minimum uncertainty only appears
perturbatively, not just in our analysis of
Schwarzschild-AdS black holes, but also in other
systems which allow a full quantization both from the
perturbative and nonperturbative points of view, casts
doubts on the believe that universal time resolution
bounds, applicable to generic quantum states,
represent an essential feature of a full quantum
theory of gravity and suggests the possibility that
they are indeed eluded when a nonperturbative
quantization of these systems is carried out.

\acknowledgments

The authors want to thank V. Aldaya and C. Barcel\'o
for fruitful conversations and enlightening
discussions. P.G. is also very thankful to F. Barbero
and J.M. Mart\'{i}n-Garc\'{i}a for their valuable
help. P.G. gratefully acknowledges the financial
support provided by the I3P framework of CSIC and the
European Social Fund. This work was supported by funds
provided by the Spanish MEC Projects No.
FIS2005-05736-C03-02 and FIS2006-26387-E.

\section*{Appendix: Calculation of Laplace integrals}
\setcounter{equation}{0}
\renewcommand{\theequation}{A.\arabic{equation}}

In this appendix we explain the computation of the
mean value of the operators $\hat{V}$ and $\hat{V}^2$
introduced in Sec. III, restricting the quantum states
to have the Gaussian behavior given in Eq.
(\ref{state}). We will obtain asymptotic expansions
for these quantities in powers of the cosmological
constant, assuming that this constant is small.

Expression (\ref{Vn}) for the mean values of powers of
$\hat{V}$ is a Laplace integral. To compute it, we
split each integral into two parts. One is integrated
from $-\gamma$ to $0$ and the other from zero to
infinity. For the first part, we make the change of
variable $v=-\sqrt{z}$, whereas for the second part we
make $v=\sqrt{z}.$ Thus, we obtain
\begin{eqnarray}
\langle\hat{V}^n\rangle&=&\frac{1}{\Xi(\mu)}
\frac{1}{\sqrt{\lambda}}\left[I_{n-}(z)+I_{n+}(z)\right],
\nonumber\\I_{n\mp}(z)&=&\int_0^{r_\mp} \frac{dz\,
e^{-z/\lambda}}{2\sqrt{z}
\left[1+(\gamma\mp\sqrt{z})^2\right]^n},
\end{eqnarray}
with $r_-=\gamma^2$ and $r_+=\infty$.

These integrals belong to a large family of the form
\begin{equation}I(z)=\int_0^r dz
F(z)e^{-\Omega z},\quad r>0.\end{equation} In the
limit $\Omega\rightarrow+\infty$, Watson's lemma
\cite{asymp} gives the whole asymptotic expansion of
any integral of this kind provided that $F(z)$ is
continuous in the interval $0\leq z\leq r$ and admits
an asymptotic series expansion of the type
\begin{equation}\label{asymp}F(z)=
z^\alpha \sum_{k=0}^{\infty}a_k z^{\beta k}, \quad
z\rightarrow0^+.\end{equation} It is necessary that
$\alpha>-1$ and $\beta>0$ for the integral to converge
at $z=0.$ In addition, if $r=+\infty,$ one must have
$F(z)\ll e^{qz}(z\rightarrow+\infty)$ for some
positive constant $q$. If the previous conditions are
fulfilled,
\begin{equation}\label{lemma}I(z)\approx
\sum_{k=0}^{\infty}\frac{a_k\Gamma(\alpha+\beta k+1)}
{\Omega^{\alpha+\beta k+1}},
\quad\Omega\rightarrow+\infty.\end{equation} Here, the
symbol $\Gamma$ represents the Gamma function
\cite{grad}.

In our case, we have in the integrands
\begin{equation}F_{n\mp}(z)=\frac{1}{2\sqrt{z}}
\frac{1}{(1+\gamma^2)^n}\left(1\mp\frac{2\gamma\sqrt{z}}
{1+\gamma^2}+\frac{z}{1+\gamma^2}\right)^{-n}.\end{equation}
If we now call
\begin{equation}x=\frac{\gamma}{\sqrt{1+\gamma^2}},\quad
y=\frac{\sqrt{z}}{\sqrt{1+\gamma^2}},\end{equation} we
realize that the factor in the parenthesis of the
previous expression is the generating function of the
Chebyshev polynomials of the second kind, $U_k(x)$
\cite{grad}:
\begin{equation}\label{Gfunmp} G(x,\mp y)=
\frac{1}{1-2x(\pm y)+y^2}=\sum_{k=0}^\infty U_k(x)(\pm
y)^k.\end{equation} Therefore, it is straightforward
to find the asymptotic series expansion of the
functions $F_{n\mp}.$ According to Watson's lemma, the
upper limit of integration $r$ does not affect the
asymptotic power series expansion of the integral. So,
we can first sum the functions $F_{n-}$ and $F_{n+}$
to obtain a unique asymptotic series for the integrand
and then apply formula (\ref{lemma}).

For $n=1$, for instance, we split the series expansion
of $G(x,\mp y)$ in even and odd powers of $y$.
Obviously, the contributions of all odd powers cancel
out when we sum $G(x,y)$ and $G(x,-y).$ We hence get
\begin{equation}F_{1-}(z)+F_{1+}(z)=
\frac{1}{\sqrt{z}}\sum_{k=0}^\infty
\frac{U_{2k}(x)}{(1+\gamma^2)^{k+1}}z^{k}.\end{equation}
By comparing this with Eq. (\ref{asymp}) we find
\begin{equation}\alpha=-\frac{1}{2},\quad \beta=1,
\quad a_k=\frac{U_{2k}(x)}
{(1+\gamma^2)^{k+1}}.\end{equation} Introducing these
values in Eq. (\ref{lemma}) with
$\Omega=\lambda^{-1}$:
\begin{equation}I_{1-}(z)+I_{1+}(z)=
\sqrt{\lambda}\sum_{k=0}^\infty\frac{\Gamma(k+1/2)}
{(1+\gamma^2)^{k+1}}\lambda^{k}
U_{2k}(x),\end{equation} where
\begin{equation}\Gamma(k+1/2)=\frac{(2k-1)!!}{2^k}
\sqrt{\pi},\end{equation} and we adopt the convention
$(-1)!!=1.$ Besides, in the sector of infinite black
hole mass ($\mu\rightarrow\infty$), we can write
$\Xi(\mu)\approx\sqrt{\pi}$ up to negligible
exponential corrections. Thus
\begin{equation}\langle\hat{V}\rangle
\approx\frac{1}{1+\gamma^2}\sum_{k=0}^\infty
\left[\frac{\lambda}{2(1+\gamma^2)}\right]^k
(2k-1)!!\,U_{2k}(x).\end{equation}

For $n=2$, repeating the above steps, Watson's lemma
leads to
\begin{eqnarray}\langle\hat{V}^2\rangle&\approx&
\frac{1}{(1+\gamma^2)^2}\sum_{p=0}^\infty
\left[\frac{\lambda}{2(1+\gamma^2)}\right]^p
(2p-1)!!\nonumber\\ &&\times\sum_{k=0}^{2p}
U_{k}(x)U_{2p-k}(x).\end{eqnarray} Finally, let us
recall that the first Chebyshev polynomials of the
second kind are $U_0(x)=1$ and $U_1(x)=2x$.


\begin{thebibliography}{99}
\bibitem{note0} We adopt units in which $\hbar=c=G=1$, with
$\hbar$ being Planck constant, $G$ Newton constant,
and $c$ the speed of light.
\bibitem{note} This implies a nonvanishing uncertainty in
any physical time interval, since we have identified
the zero time with the instant at which we start the
measurement process.
\bibitem{nota} We consider
states with $\langle\hat{M}\rangle < \infty$ and, if
needed, take on them the limit $\Delta M\rightarrow
0$.
\bibitem{nota2} It suffices that there exists a
positive constant $C$ such that
$\langle\hat{M}^{2n}\rangle<C^n$ $\forall n\geq 1$.
\bibitem{note2} For this, we
have assumed that the time $\overline{T}$ can be large
and restricted ourselves to states with bounded
expectation values $\langle\hat{M}^{2n}\rangle$ as
explained in \cite{nota2}.
\end{thebibliography}
\end{document}